\documentclass[12pt,preprint]{aastex}

\slugcomment{PASP in press}
\shorttitle{Brown dwarfs in eclipsing binary systems}
\shortauthors{Pinfield et al.}

\begin{document}

\title{The prospects for finding brown dwarfs in eclipsing binary systems and measuring brown dwarf properties}

\author{D. J. Pinfield\altaffilmark{1}, H. R. A. Jones\altaffilmark{1} and 
I. A. Steele\altaffilmark{1}}
\affil{Astrophysics Research Institute, Liverpool John Moores University, 12 Quays House,\\
Egerton Wharf, Birkenhead, CH41 1LD}
\email{dpi@astro.livjm.ac.uk}

\begin{abstract}

We present the results of a simulation to investigate the prospects of measuring 
mass, age, radius, metallicity and luminosity data for brown dwarfs in fully eclipsing 
binary systems around dwarf spectral types from late K to early M that could be 
identified by ultra-wide-field transit surveys such as SuperWASP. These surveys 
will monitor approximately a million K and M dwarfs with $|b|>20^{\circ}$ (where blending 
of sources is not a significant problem) at a level sufficient to detect transits of 
low luminosity companions. We look at the current observational evidence for such 
systems, and suggest that $\sim$1\% of late K and early-mid M dwarfs could have a 
very close ($\sim$0.02AU) BD companion. With this assumption, and using SuperWASP 
as an example, our simulation predicts that $\sim$400 brown dwarfs in fully eclipsing 
binary systems could be discovered. All of these eclipsing binaries could yield 
accurate brown dwarf mass and radius measurements from radial velocity and follow-up 
light curve measurements. By inferring the brown dwarf effective temperature distribution, 
assuming a uniform age spread and an $\alpha$=0.5 companion brown dwarf mass function, 
the simulation estimates that brown dwarf brightness could also be measurable (at the 
10\% level) for $\sim$60 of these binary systems from near infrared follow-up light 
curves of the secondary eclipse. We consider irradiation of these brown dwarfs by 
their primary stars, and conclude that it would be below the 10\% level for $\sim$70\% 
of them. This means that in these cases the measured brown dwarf brightnesses should 
essentially be the same as those of free-floating counterparts. The predicted age 
distribution of the primaries is dominated by young systems, and $\sim$20 binaries 
could be younger than 1Gyr. Irradiation will be below the 10\% level for $\sim$80\% 
of these. We suggest that many of these young binary systems will be members of 
``kinematic moving groups'', allowing their ages to be accurately constrained.
\end{abstract}


\section{Introduction}
The mass of a brown dwarf (BD) is arguably its most basic characteristic. The essence of 
the classification of an object as a BD depends on the objects sustaining deuterium burning 
(for solar metallicity $>$0.013M$_{\odot}$) but not sustaining hydrogen burning (for solar 
metallicity $<$0.075M$_{\odot}$). However, since BDs contract, cool and fade with time, their 
observable properties depend strongly on both age and mass. Furthermore, because BD colours 
are strongly influenced by atmospheric dust formation and molecular absorbers (like H$_2$O), 
their appearance will depend not only on effective temperature (T$_{eff}$) and surface gravity 
($g$), but also on composition. Accurate BD mass, age, radius, metallicity and luminosity data 
are thus vital to properly test our understanding of these extreme objects.

In view of the particular importance of such data for BDs, it is unfortunate that it has thus 
far not been possible to completely characterise {\it any} BDs in this way. There are several 
reasons for this. In order to measure BD masses, one must find BDs in binary systems, and measure 
the orbits. One also needs to be able to measure the BD itself to determine its luminosity, and 
the age of the binary must be established, so as to infer the age of the BD. Proper motion 
measurements have shown that $\sim$10\% of solar type stars have BD companions at wide ($>$1000AU) 
separations (Gizis et al. 2001). However, the periods of such binaries are much too long to be 
useful in measuring masses. A small number of possible close BD companions to solar type stars 
have been identified (Halbwachs et al. 2000; Nidever et al. 2002; Blundell et al. 2004). But such 
systems are very rare, and could not be directly imaged anyway because of the glare of the primary. 
For the lowest mass stars, adaptive optics imaging has shown that close ($\sim$1AU) BD companions 
to late M and early L dwarfs are significantly more common (Close et al. 2003), with orbits that 
can be measured in a few years, and binary components that can be individually studied. For example, 
Bouy et al. (2004) astrometrically measured 36\% of the 10 yr orbit of 2MASS 0746+2000AB, constraining 
the system mass at the 10\% level. However, the ages of such binaries will in general be very difficult 
to constrain without recourse to the theoretical models that one wishes to test. For instance, Bouy et~al. 
compared 2MASS 0746+2000 A and B to Dusty model predictions (Chabrier et~al. 2000) in a K, J-K colour-magnitude 
diagram (CMD) and estimated a system age of 150--500Myrs. However, when compared to late M 
and L dwarfs with accurate parallaxes, 2MASS 0746+2000 A and B do not lie above the CMD sequence 
as sources of this age should. A potentially more useful type of system is represented by the 
Epsilon Indi multiple, which consists of a double BD binary (visual separation of 2.65AU) at a 
distance of 1500AU from its K5 primary (McCaughrean et al. 2004). The BD binary orbit should be 
measurable (P$\sim$15yrs), and the age of the system can be constrained to some degree (0.8-2Gyrs) 
by measurements of the K dwarf primary (Lachaume et al. 1999). However, such multiple systems are 
not common place, and age constraints established from measurements of a single primary star will 
not generally be very precise.

The concept presented here is to exploit the new breed of ultra-wide-field transit surveys (designed 
primarily to find signatures of planetary transits) to identify BDs in fully eclipsing close binary 
systems around cool stars. BD masses could be measured from the doppler wobble of the primary (using 
radial velocity techniques). One could measure BD radii from the depth of the primary eclipse (when 
the BD crosses the face of the primary). The brightness of the BD may be measured from the secondary 
eclipse (when the BD goes behind the primary) provided the BD is bright enough compared to the primary. 
Many of these systems will be young (because BDs are brighter when young), making membership of young 
kinematic moving groups more likely. Membership of such moving groups would accurately constrain 
age and metallicity.

In this paper, we discuss the prospects of identifying BDs in binaries capable of providing the 
full compliment of BD mass, age, radius, metallicity and luminosity data. 
In Section 2 we discuss what is currently known about the frequency and separation of close BD 
companions, and how this effects the likelihood of eclipse. In Section 3 we use the SuperWASP 
survey as an example of the new ultra-wide-field transit surveys, estimate sensitivities and the 
effects of source crowding in images with very large pixels, and estimate the distances out to 
which BD transits should be detectable. Section 4 describes the additional factors limiting the 
number of suitable binary systems; the accuracy with which secondary eclipses can be measured 
(Section 4.1), and the limits imposed by the doppler wobble technique (Section 4.2). Section 5 
gives some background on kinematic moving groups, and details how one could accurately constrain 
the ages of young systems. Section 6 describes our simulation to predict the size and properties 
of the eclipsing BD population. Section 7 then discusses some identification and followup 
procedures, and Section 8 presents our conclusions.

\section{Close BD companions}

\subsection{Frequency and separation}
In Table 1 we show and compare what is known about the frequency of companion stars and BDs 
as a function of separation. For solar type stars, Halbwachs et al. (2003) and Duquennoy \& 
Mayor (1991) have measured the frequency of stellar companions as a function of period (which 
we convert into separation), using 10 years of radial velocity data combined with visual 
binaries and common proper motion systems. It is clear from this work that stellar companions 
are common over a large range of separation. However, it has been known for some time that 
there is a dearth of low mass ratio (q$\le$0.1) companions in close orbits around solar type 
stars, known as the ``BD desert''. Other radial velocity surveys (eg. Marcy \& Benitz 1989; 
Fischer \& Marcy 1992; Nidever et al. 2002) have now studied large numbers of solar type stars, 
and it is currently estimated that $\le$1/500 have a BD companion within 3AU (Vogt et al. 2002). 
This corresponds to a BD companion fraction of $<$0.07\% per decade in separation (clearly much 
less than the $\sim$4\% for stellar companions. The desert does not extend to wide separations, 
with sub-stellar companions imaged at 15--20AU (Els et~al. 2001; Lui et al. 2002), and BDs 
are as common in very wide binaries ($\sim$1000AU or more) as stellar companions around solar 
type stars.

In terms of primary type, there is evidence (from the OGLE transit survey) that the desert 
extends out to the mid-K dwarfs, since the majority of the small low luminosity companions 
discovered by Udalski et al. (2002a, b, 2003) turned out to be low-mass stars and not BDs (Dreizler 
et al. 2002). However, for later dwarfs, the target lists for radial velocity surveys are not large 
enough to accurately constrain the close BD companion fraction. Despite these limitations some 
interesting candidate BD companions have been discovered. Reid \& Mahoney (2000) found a 
possible BD companion (0.07--0.095M$_{\odot}$) at a separation of $\sim$0.005--0.007AU from 
an $\sim$M5 dwarf in a radial velocity survey of $\sim$50 Hyades M dwarfs. Nidever et al. 
(2002) found 2 possible BD companions (at separations of $\sim$0.2 and $\sim$2AU) around early 
M dwarfs from a radial velocity survey including $\sim$100 M dwarfs. Conversely, Marchal (2003) 
found no close BD companions in a comparable radial velocity survey of $\sim$100 M0-5 dwarfs.

At somewhat larger separations, Oppenheimer et al. (2001) found 1 BD in a coronagraphic survey of the 
stellar population within 8 parsecs of the Sun. This volume limited sample (ie. mostly 
M dwarfs) consisted of 163 stars in 111 star systems, and BDs with mass$\ge$0.04M$_{\odot}$ 
and age $\le$5Gyrs could have been detected with separations of 40--120AU around 80\% of the 
sample. BDs are not found in significantly wider orbits around early M dwarfs, perhaps because 
impulsive perturbations by close stellar encounters will cause such weakly bound systems to 
dissolve (see Burgasser et al. 2003). It is not, then, surprising that this lack of wide BD 
companions is also seen around the latest stars and BDs (with the possible exception of one 
very young candidate BD binary which would have a separation of $\sim$240AU; Luhman 2004). 
However, closer companions to late M, L and T dwarfs are quite common (see Table 2), and while 
accurate masses are generally not available, many of these will be sub-stellar (eg. Freed et al. 
2003; an L7.5 BD $\sim$3AU from an M8 star).

Of interest here are the late K--mid M dwarf primaries, since we will subsequently 
show that sufficient numbers of these dwarfs will be monitored by ultra-wide-field surveys to 
find eclipsing BD companions (see Section 6). It is clear from Table 1 that BD companions to 
these stars could occur at the $\sim$2\% level in orbits close enough to make transits likely 
($\le$0.02AU; see Section 2.2). However, an accurate determination of the masses of the BD 
candidates discovered so far would require knowledge of orbital inclination. Some may turn out 
to be low-mass stars at low inclination (eg. Halbwachs et al. 2000). If correct though, it would 
show that although the BD companion fraction in close orbits around such stars is lower than 
that of stellar companions, there would be no BD desert for these primaries.

In addition to these systematic searches, there have been two serendipitous discoveries of 
possible close BD companions. Schuh et al. (2003) reported the discovery of an eclipsing BD 
companion to a late K dwarf with a separation of 0.02AU. This system was discovered while 
photometrically monitoring a white dwarf. Also Santos et al. (2002) discovered a radial 
velocity BD companion to an M2 dwarf with a separation of 0.017AU. This system was found 
as a result of the weak gravitational wobble it induced in the visual K2 companion of the 
M dwarf.

Having considered all known candidates, it can be seen that the spectral type range of the 
primaries in systems that could potentially harbour significant populations of very close BD 
companions ranges from late K to mid M. The orbital periods of these BD companions in close 
orbits around late K--mid M dwarfs is 1--1.5 days. At these separations, a full transit lasts 
$\sim$1.5hrs.

\subsection{Eclipsing fraction}
The majority of the known candidate BD companions around K and M dwarfs (three out of five) are in close 
orbits with separations $\le$0.02AU (see previous section). Close systems such as these will be 
the most likely to eclipse, because eclipsing probability is $\propto 1/a$ (where $a$ is separation). We 
therefore chose to consider the likelihood of an eclipse for a BD companion at a separation of 0.02AU. 
This separation corresponds to $\sim$10R$_{\star}$ for an early M dwarf, and for such separations, a 
favourable $\sim$1 in 15 should be eclipsing. The requirement for a full eclipse is that 
\begin{equation}
\sin{(90-i)} \le \frac{R_p - R_{BD}}{a},
\end{equation}
where $i$ is the inclination angle, $R_p$ and $R_{BD}$ are the primary and BD radii respectively, and 
$a$ the component separation. The fraction of sources that should be eclipsing is $(90-i)/90$. 
Using theoretical stellar radii from the 500Myr NextGen models of Baraffe et al. (1998), and assuming 
0.1R$_{\odot}$ for the BDs (typical for 500Myr age), we get the eclipse fractions shown in Figure 1. 
As the primary radius decreases, the chances of a full eclipse fall off, and for comparable radii 
full eclipses are very unlikely. It can be seen that for late K--early M dwarfs, the eclipsing fraction 
is $\sim$4-7\%, and so even with a $\sim$1\% binary fraction at 0.02AU, one could still expect to find 
fully eclipsing BD companions around $\sim$1 in 2000 dwarfs of this type.

\subsection{Irradiation}
Since we are interested in measuring companion BD properties that would be essentially the same as 
those of free-floating counterparts, we desire irradiation effects to be low. However, for the very 
close BD companions discussed in Section 2.2, this may not always be the case, since the cool BD companion 
could be heated by the nearby brighter hotter primary, as happens with giant planets in close orbits 
around solar type stars (hot Jupiters). Hot Jupiters are strongly irradiated, and have higher T$_{eff}$s 
and swollen radii as a result (eg. Burrows Sudarsky \& Hubbard 2003; Chabrier et al. 2004). Irradiation 
should only be an issue when the incident flux from the primary is significant compared to the emergent 
flux from the companion. If the incident flux is 10\% of the emergent flux for instance, then the BD 
luminosity would increase by no more than 10\% (and the radius and T$_{eff}$ by much less than this). 
This level represents a useful level of accuracy with which to measure BD brightness. Figure 2 shows 
how the amount of irradiation varies for BD companions (spectral types M6-T4) at 0.02AU separation 
from their primary stars. Each point in the plotting area represents a binary where the primary and 
secondary spectral types can be read off the y and x axis respectively. Using NextGen model data for 
the primaries, BD T$_{eff}$s appropriate to their spectral type (Reid et al. 1999; Knapp 2004) and 
assumed BD radii of 0.1R$_{\odot}$, we derived a locus of binary systems in which the BD will be 
irradiated by 10\% of its emergent flux. This locus is plotted in the figure as a solid line, where 
binaries above the line are not significantly irradiated. For example, an M8 BD will not be significantly 
irradiated at 0.02AU from a mid K dwarf (or later) primary, and an L4 BD will not be significantly 
irradiated at 0.02AU from an M0-2 (or later) primary.

\subsection{Age and mass distributions}
The mass and age distribution of the companion BDs is important because the older and lower mass 
BDs will be more difficult to measure from the secondary eclipse. The age distribution should be 
the same as the local disk. An analysis of chromospheric activity of G and M dwarfs in the solar 
vicinity (Soderblom, Duncan \& Johnson 1991; Henry, Soderblom \& Donahue 1996) is consistent with 
a uniform star formation rate over the age of the disk, and we thus assume that BDs will have a uniform 
age distribution between 0 and 10Gyrs. The mass distribution of companion BDs is less constrained. 
The BD mass function slope for the field and young clusters/associations ranges from $\alpha\sim$0.5-1 
(where $dn/dm\propto m^{-\alpha}$). However, the masses of BDs in close binaries may have a very 
different distribution, since dynamical factors during formation (eg. Reipurth \& Clarke 2001) and 
possible radial migration (due to early disk interaction; Armitage and Bonnell 2002) could be important. 
For very low-mass binary systems for instance, the q distribution peaks at 0.7 -- 1 (Burgasser et~al. 
2003; Pinfield et~al. 2003). Such a bias amongst binaries with an early M dwarf primary would produce 
a strong preference for stellar companions over BDs, and the BD companion mass function itself would 
prefer high mass BDs. However, it is not currently possible to estimate $\alpha$ for close BD companions 
from the limited observational evidence, and it essentially remains a free parameter. Low values of 
$\alpha$ would result in proportionally more high mass BDs (for some value of the BD companion fraction), 
which would be brighter and easier to detect. Higher values would mean proportionally more low-mass BDs, 
many of which would be too faint to detect.

\section{Ultra-wide-field transit surveys}
The discovery of the first transiting extra-solar planet (Charbonneau et al. 2000; Henry et al. 2000) 
has led to a proliferation of ground based photometric searches aiming to detect planets around bright 
solar type stars by searching for transit signatures in photometric light curves (see Horne 2003). 
These searches fall into two categories; the deeper, but relatively small area searches such as 
the Optical Gravitational Lensing Experiment (which uses a 1.3m aperture telescope), and the 
ultra-wide-field surveys such as SuperWASP (Christian et al. 2004), TrES (Alonso et al. 2004), 
Vulcan South (Caldwell et al. 2003) and others, that use small aperture ($\sim$10 cm) wide-field 
(5--10$^{\circ}$) CCD based systems with spatial resolution of $\sim$10'' pixel$^{-1}$. While 
accurate light curve extraction is more challenging for ultra-wide-field data, the ultra-wide-field 
surveys promise to discover transits around brighter nearer stars, for which close transiting 
companions can be more accurately measured. In this paper we focus on the abilities of these 
ultra-wide-field surveys, and use the UK's SuperWASP project as a specific example.
\subsection{SuperWASP}
SuperWASP (Super Wide Angle Search for Planets) is a project to construct and operate 
facilities to carry out ultra-wide-field unfiltered photometric monitoring surveys (see Pollacco 
et al, www.superwasp.org). The first SuperWASP facility (SuperWASP-I) is located on Roque de Los 
Muchachos, La Palma, and a second facility has also been funded for the southern hemisphere. 
SuperWASP-I consists of five camera units mounted on a large computer controlled fork mounting, 
housed in a small enclosure. Each camera unit consists of a 200mm f1.8 Canon lens with an aperture 
of 11.1cm (made of fluorite and UD glass components) and a 2048x2048 pixel e2v42 thinned CCD (BV 
coated) cooled to -60C by a 3-stage Peltier. Each camera has an 8$^{\circ}$x8$^{\circ}$ field of 
view, resulting in large 14.3 arcsecond CCD pixels. The combined sky coverage of the five cameras 
is thus $\sim$320 sq. degs. SuperWASP-I current observing strategy is to repeatedly measure 30s 
integration images of the same region of sky throughout each night for a period of $\sim$2 months, 
before moving onto a new region. With low overheads ($\sim$4s readout), SuperWASP-I takes $\sim$100 
sets of images per hour. First light was in November 2003, and at time of writing SuperWASP-I is 
routinely taking data. A data reduction pipeline is being implemented, and light curves of all 
stars measured will be placed on the public ``Ledas database'' at The University of Leicester, 
where software search algorithms will be available for data mining.

\subsection{SuperWASP sensitivities}
SuperWASP is an unfiltered system, and as such the transmission profile (or pass band) is determined 
by the atmosphere, the optical throughput of the lenses and CCD entrance window, and the CCD 
quantum efficiency. To estimate the shape of the SuperWASP transmission profile we have used available 
site and technical details. We used La Palma atmospheric extinctions from the ``INT 
Wide Field Survey'' web-site (based on observations by Derek Jones in August 1993). For the 
Canon lenses, the optical transmission of UD glass and fluorite ($\sim$95\% per pair of air-glass 
interfaces) does not vary significantly from $\sim$0.23-7 microns (minimising chromospheric 
aberration). We thus assumed a wavelength independent optical transmission of 75\% to allow 
for the lens groups (in each Canon lens) and the CCD entrance window. The CCD quantum 
efficiency was taken from the E2V CCD selection guide, interpolating between the -20$^{\circ}$C 
and -100$^{\circ}$C curves. Figure 3 shows the atmospheric, lens and CCD transmission profiles 
(dotted lines) against wavelength, and the resulting total SuperWASP transmission (solid line). The 
transmission profile peaks in the V and R bands (at $\sim$60\%). Blue of V, both atmospheric 
extinction and the CCD quantum efficiency reduce throughput. Red of R the transmission is 
predominantly reduced by CCD quantum efficiency. We estimated a system zero point using a 
spectrum of Vega (from Hayes, 1985 and Mountain et al., 1985), which we take to represent 
a zero magnitude source. Accounting for the collecting area (96.7 sq. cm for each lens) 
and the SuperWASP transmission, we integrated the detected Vega photons across the SuperWASP 
band. The predicted zero point was found to be 21 magnitudes (2.5x10$^8$ detected counts/s from a 
zero magnitude source). As a guide, this means that a signal-to-noise of $\sim$50 would result 
from a 30s integration on a 16$^{th}$ magnitude source.

\subsection{Unfiltered colour term}
We used the spectra of Pickles (1985; available on CDS) and Leggett et al. (2000; from 
www.jach.hawaii.edu/$\sim$skl), combined with a Johnson V-band (Bessell 1990) and the SuperWASP 
pass band, to derived synthetic V-band and SuperWASP-band flux ratios for early G-mid M dwarfs. 
These flux ratios were converted into SuperWASP colours (V-m$_{SW}$, where m$_{SW}$ is a magnitude 
on the natural SuperWASP system) using our Vega spectrum to define zero colour. Figure 4 shows 
the resulting colour terms as a function of spectral type. It can be seen that for early G dwarfs 
the colour term is small, but for later spectral types it becomes significant because most of the 
optical flux from these dwarfs is not in the V-band, but is in the red part of the SuperWASP band.
\subsection{Unfiltered sky brightness}
To estimate the La Palma sky brightness as seen by SuperWASP, we used the 
full range of optical sky brightness measurements from Benn \& Ellison (1998), for bright, 
grey and dark sky conditions. For each sky we converted the magnitudes into fluxes using 
our Vega spectrum (see Section 3.1). The resulting flux levels were then joined together 
to provide an approximate sky spectrum. These sky spectra were then multiplied through by 
the SuperWASP band pass, and converted into SuperWASP magnitudes using our system zero point. 
The SuperWASP sky brightness estimates were found to be 18.3, 20.3 and 21.1 magnitudes per square 
arcsecond for bright, grey and dark skies respectively. If one accounts for the large SuperWASP 
pixels and a spatial resolution element containing 12 pixels (a 2 pixel radius aperture), one 
would expect sky brightnesses of 9.8, 11.8 and 12.6 magnitudes per resolution element, for 
bright, grey and dark skies respectively.

\subsection{Source crowding with large pixels}
The large SuperWASP pixels (and consequent large aperture sizes used for photometry) mean 
that image crowding will be much more of an issue than it is for higher spatial resolution 
imaging, particularly in the Galactic plane. This is important for two reasons. Firstly, if a 
transit source shares its aperture with another star (ie. if it is blended), the SNR of the 
transit detection will be decreased. However, more importantly, if the PSF of a neighbouring 
source spills into the aperture of a star, this fraction of the measured flux could vary over 
time if the PSF changes slightly, or the chosen aperture does not remain perfectly centred on 
the star -- SuperWASP tracking ($\sim$0.01arcsecond/s accuracy) will cause stars to drift 
between pixels over several hours. Since we are interested in identifying eclipse dips that 
are $\sim$4\% deep, we would not want such flux pollution much above the 1\% level. We have 
therefore used synthetic gaussian point-spread-functions to determined the distance at which 
a source (with FWHM=1.5 pixels) will pollute it's neighbours 2 pixel aperture by 1\%, and 
established how this distance ($d_{1\%}$) depends on the relative brightness of the two sources 
($\Delta mag$). We found that 
\begin{equation}
  d_{1\%} = FWHM \times ( 1.95 - 0.12\Delta mag ),
\end{equation}
for $\Delta mag <$2.0. For fainter neighbours, $d_{1\%}$ drops more rapidly, reaching zero 
for $\Delta mag=$5 (ie. a factor of 100). So if a source has a neighbour with $d<d_{1\%}$, we 
consider it significantly blended, and not measured with sufficient accuracy for our purposes 
by SuperWASP. We applied this criteria to a series of 1 sq. deg catalogues (centred on different 
Galactic coordinates) from the SuperCOSMOS Sky Survey database\footnote{http://www-wfau.roe.ac.uk/sss/}. 
Since we are interested in late K and early M dwarfs, we only considered SuperCOSMOS sources with 
B$_J$-R$\ge$1.8, and estimated the fraction of blends as a function of Galactic coordinate and 
source brightness. We found that for V$\sim$16, $\sim$30\% of sources with $|b|>20^{\circ}$, and 
$\sim$75\% of sources with $|b|<20^{\circ}$ will be blended in SuperWASP images. For V$\sim$18, 
we found 50\% and 85\% blended fractions for $|b|>$ and $|b|<20^{\circ}$ respectively.

\subsection{Transit detection magnitude limits}
The signal-to-noise level of a transit detection is 
\begin{equation}
SNR = \left( \frac{R_{BD}}{R_{\star}}^2 \right) 
      \left( \frac{S_{\rm obs}}{noise_{\rm obs}} \right) 
      \sqrt{\frac{P R_{\star}}{t_{\rm obs} \pi a}},
\end{equation}
where $R_{BD}$ and $R_{\star}$ are the radii of the companion BD and the primary star, 
$S_{\rm obs}$ and $noise_{\rm obs}$ are the signal detected from the source and the associated 
noise during an observation, $P$ is orbital period, $t_{\rm obs}$ is the time taken for an 
observation ($\sim$30s), and $a$ is the separation of the binary. The first term represents 
the relative decrease in brightness during transit, the second term is the signal-to-noise 
associated with each observation, and the last term is the square root of the number 
of observations made during a transit. The noise model we used was 
\begin{equation}
noise_{\rm obs} = \sqrt{ (S_{\rm obs}t_{\rm obs}) + (S_{\rm sky}t_{\rm obs}) + 
(D t_{\rm obs}) + (N R^2) + (Syst^2) },
\end{equation}
where $S_{\rm sky}$ is the signal detected from the sky in an $N$=12 (2 pixel radius) spatial 
resolution element (cf. the lenses produce a $\sim$1.5 pixel point-spread-function) during an 
observation, $D$ is the dark current ($\sim$0.01$e^-$/s), $R$ is the read-noise ($\sim$12$e^-$), 
and $Syst$ represents any systematic errors. In general, systematic uncertainties associated with 
relative photometry of bright sources can be reduced to the 1--2 m-mag level (Kane et al. 2004; Bakos 
et al. 2004). For the fainter sources, of interest here, systematics will be dominated by field 
crowding effects which sets the upper level of the systematic uncertainties at a self imposed 10 
m-mags (see Section 3.4).

We define our transit detection limits requiring a SNR$\ge$5. We would only expect 1 in $\sim$2 
million stars to register a 5-$\sigma$ transit due to random noise alone, and since SuperWASP will 
survey rather less than this number of late K and M dwarfs, false positives from random noise will 
not be a problem. We combined equations 2 and 3 with the SuperWASP zero point, and used the binary 
characteristics and stellar model data described in Section 2.2 to derive SuperWASP magnitude limits 
for 5-$\sigma$ transit detection. These limits, for bright, grey and dark skies are shown in Figure 
5 (solid lines), as a function of primary M$_V$. Primary spectral types are also indicated at their 
appropriate position (taken from Pickles 1985). V-band magnitude limits were then derived using the 
colour term from Section 3.2 (dashed lines). For comparison, the G star 10-$\sigma$ bright sky 
detection limit of Horne (2003) is over-plotted (open circle), and joined to its 5-$\sigma$ equivalent 
(filled circle). It is clear from the figure that as one moves to later type stars, transits can be 
detected out to fainter magnitudes. This is because later stars are smaller, and eclipse dips 
are correspondingly deeper. Also, the effect of the broad unfiltered pass band is clear, since 
SuperWASP will do better for red sources than a comparable V-band filtered system.
%

\subsection{Distance limits for transit detections}
The V-band transit detection limits from Figure 5 were converted directly into distance limits 
using the appropriate values of M$_V$. These distance limits are shown in Figure 6 (solid lines) 
for bright, grey and dark skies. It can be seen that transits of late K-early M dwarfs should be 
detectable out to $\sim$200-500pc. For the later mid M dwarfs, SuperWASP is limited to $\sim$100pc, due 
to their fainter intrinsic brightness. The figure also contains some additional limits, which 
will be described in the next section.

\section{Limits imposed by follow up requirements}

\subsection{Measuring BD brightness from secondary eclipses}
Having identified a good transit signature of a low luminosity fully eclipsing companion 
(and confirmed it with an intermediate sized optical telescope) , one can then attempt 
to measure the brightness of the companion from the secondary eclipse. Our ability to do this 
is governed by the accuracy with which we can measure follow up light curves and the relative 
brightness of the companion and primary. Compared to late K/early M dwarfs, BDs with spectral 
types $<$T2 are relatively brightest in the K-band (see Leggett et al. 2002). K is generally 
thus the best band in which to search for a secondary eclipse. With M$_V$=8--12 and M$_K$=5--7 
(see Reid \& Cruz 2002), a typical late K/early M dwarf near the SuperWASP distance limits 
(500-200pc respectively) would have K$\simeq$13.5. The accuracy of K-band light curves will 
depend on photon noise, stability of the instrumental response, and atmospheric conditions. 
Using 2--4m class telescopes, a signal-to-photon-noise ratio of $\sim$1000 could be achievable 
for a K$\sim$13.5 source in $\sim$5-10 minutes. Variations in the zero point and extinction 
can be corrected for using frequent standard measurements, flat-fielding should be good to at 
least 0.1\% accuracy, and atmospheric scintillation noise for such observations will be $\sim$1 
part in 10$^4$ (equation 10 of Dravins et~al. 1998). The main source of systematic uncertainty 
in near infrared photometry is the amount of atmospheric H$_2$O absorption, which can be variable 
on very short timescales. Kidger \& Mart\'{i}n-Luis (2003) measured high precision near infrared 
photometry of bright (K$\le$10) northern hemisphere stars using a photometer, and were able to 
obtain K-band accuracies (inferred from multiple measurements) as good as 0.4m-mags for some 
sources (see their Figure 8). Typical uncertainties would not always be this good, as atmospheric 
conditions change (cf. Kidger \& Mart\'{i}n-Luis' median uncertainty was $\sim$3m-mags), however 
an overall K-band measurement accuracy of 1m-mag should be a realistic level to aim for. Note 
that large scale NIR arrays (such as WFCAM on the UK IR Telescope) would allow comparison stars 
to be measured at the same time as targets, which could reduce atmospheric systematics. However, 
we will continue to use Kidger \& Mart\'{i}n-Luis as our main example, since they have 
observationally established these levels of accuracy.

The total accuracy with which we can measure the depth of a secondary eclipse dip in a near 
infrared light curve ($\sigma_{tot}$) will depend on both the accuracy of an individual observation 
($\sigma_{obs}$) and on the number of data points ($N$) measured during secondary eclipse (where 
$\sigma_{tot}=\sigma_{obs}\sqrt{N}$; see Henry et al. 2000). For the binaries we are considering, 
we would expect an eclipse to last $\sim$1.5 hrs, and we could thus expect to obtain 6-10 K-band 
measurements per eclipse (allowing for 5 minutes overhead per observation). By measuring 3-4 eclipses, 
and phasing the data appropriately, one could then measure the depth of the secondary eclipse dip 
with an accuracy of 0.2m-mags. Thus, for an eclipse dip to be measured with 10\% accuracy, it must 
be at least 2m-mags deep. The BD must therefore be no less than 500 times (6.75 mags) fainter than 
the primary star. Estimating primary M$_K$ from Reid \& Cruz (2002), and BD M$_K$ (appropriate for 
M6-T4 spectral types) from Dahn et al. (2002) and Knapp et al. (2004), we have derived M$_V$ brightness 
limits for primaries with M$_K$ 6.75 magnitudes brighter than BD companions of each spectral type. 
These primary M$_V$ limits are shown in Figure 6, where each is labelled with its appropriate BD 
spectral type. As an example, it should be possible to measure the secondary eclipse of an L6 
eclipsing BD companion of an M$_V\ge$9 primary.

\subsection{Measuring BD masses from radial velocities}
If a low luminosity eclipsing companion is detectable from its secondary eclipse, then radial 
velocity techniques can be employed to accurately measure its mass. For eclipsing systems, the 
inclination is known to better than 1$^{\circ}$, and the relative mass of the BD (compared to that 
of the primary) can be determined by measuring a radial velocity curve for the primary. Primary 
masses will thus act as anchors for the BD masses, and uncertainties in the mass-luminosity relation 
for low-mass stars (cf. Henry \& McCarthy 1993) will propagate through to the BD masses. However, it 
is expected that the low-mass star mass-luminosity relation should improve significantly in the next 
few years, as the orbits of many more M dwarf systems are measured (eg. by the RECONS; Jao et al. 2003).

The radial velocity accuracy required to determine BD masses places its own limits on the distance 
of useful binary systems. To illustrate this, a 0.05M$_{\odot}$ brown dwarf at 0.02AU will cause 
a radial velocity oscillation of $\pm$20kms$^{-1}$ for a 0.5M$_{\odot}$ M dwarf primary. Standard 
arc lamp echelle spectroscopy is limited by several factors; zero point drifts in the wavelength 
scale between exposures, the spectral features available for cross correlation techniques, the 
presence of telluric absorption, the signal-to-noise of the spectra themselves, and intrinsic 
properties of the star (surface spots and/or rotation). We will return to the issue of spots and 
rotation in Section 7.3. For the other factors however, we will use previous optical studies with 
8m telescopes as a guide. Reid \& Mahoney (2000) observed zero point drifts of $\pm$1kms$^{-1}$ 
during a full night for the Keck HIRES echelle spectrograph. However, they managed to obtain 
$\pm$300ms$^{-1}$ accuracy by cross correlating echelle orders (containing strong telluric 
signatures), with white dwarf spectra (which act essentially as continuum sources onto which 
a clear telluric signature is imprinted). More recently, Konacki et~al. (2003) used an iodine 
cell to more rapidly measure zero point drifts of $\sim$200ms$^{-1}$ over timescales of $\sim$30 
minutes. These studies looked mostly at K and M dwarfs from 3850--6200\AA\ and 6350--8730\AA\ 
respectively, and found that such drifts generally dominate the overall uncertainty, with 
contributions of only $\pm$60ms$^{-1}$ coming from the signal-to-noise ($\sim$15--25) of the 
spectra themselves.

It is thus clear that optical echelle spectroscopy should be capable of providing radial 
velocities with accuracies of $\pm$0.5kms$^{-1}$ (enough to constrain close BD companion 
masses at the $\sim$2--3\% level) using lower signal-to-noise spectra from rather fainter 
sources with V$\sim$19. We have converted this magnitude limit into a distance limit which 
is shown in Figure 6 (dashed line). It can be seen that for sources with spectral types 
$<$M4 the echelle followup requirements do not impose any additional distance constraints 
beyond those of SuperWASP itself. The efficient follow up of later type primaries at the 
SuperWASP distance limit would require a near infrared echelle spectrograph (such as PHOENIX 
on Gemini). However, we do not expect to find many of these as fully 
eclipsing systems (see Sections 2.2 and 6).

\section{Determining accurate ages}
\subsection{Open clusters}

Disk stars with the best constrained ages are those in open clusters. The 
age of an open cluster can be well constrained using a variety of methods, such as fitting 
the upper main sequence turn off (Sarajedini et al.1999), or (for young clusters) measuring 
the magnitude of the lithium depletion edge (Stauffer, Schultz \& Kirkpatrick 1998). An open 
cluster is believed to be a coeval association (with only a relatively small age spread 
of perhaps a few Myrs). Therefore, if membership of a cluster is established, then the 
cluster age can be assumed for the star. However, the majority of disk stars are not open 
cluster members. Within the SuperWASP distance limits ($\sim$500pc) there are $\sim$31 open 
clusters in the La Palma sky (see the LEDAS database\footnote{http://ledas-www.star.le.ac.uk}). 
Although catalogued membership of these clusters is by no means complete, a reasonable estimate 
would be that between them they contain no more than a few thousand late K/early M dwarfs. 
This represents a tiny fraction of the number of non cluster disk stars of this type in the 
SuperWASP sky ($\sim$1 million). Open cluster membership will thus not be very useful for 
constraining the age of a significant fraction of the binary systems under consideration.

\subsection{Young moving groups}

Open clusters are thought to represent only small components of larger kinematically distinct 
coeval populations that are spatially dispersed within the disk. Such disk populations are known as 
moving groups. Moving groups were originally identified by Eggen in the 1950s. However, it has since 
become clear that many of the older ($>$1Gyr) so called moving groups he identified show a spread in 
both age and metallicity (Nordstrom et al. 2004), and these kinematic signatures are more likely the 
result of disk heating by stochastic spiral waves (eg. De Simone, Wu \& Tremaine 2004). There is 
evidence however, that the younger moving groups are coeval. Castro, Porto de Mello \& de Silva (1999) 
found identical Cu and Ba abundances for 7 G and K dwarfs in the Ursa Major moving group, and King 
et al. (2003) found clear evidence of a correlation between membership of the Ursa Major moving group 
and activity levels (from CaII H and K emission strength).

Such young moving groups are thought to originate in the same environments as open clusters. Very 
young clusters should virialise somewhat within their natal gas, and when the gas is cleared by OB 
star action and the cluster expands, the stars with high enough velocities will become unbound. These 
unbound stars then slowly expand forming the moving group, and possibly leaving behind an open 
cluster consisting of any remaining bound stars (see the N-body simulations of Kroupa Aarseth \& 
Hurley 2001). Typical expansion velocities for the unbound population will be $\sim$5-10kms$^{-1}$ 
(5-10pc/Myr), and the moving group space motion will thus remain quite well defined (superimposed 
on the more dispersed disk population) until it has had time to become significantly phase mixed by 
disk heating. Models suggest that phase mixing will increase a moving group velocity dispersion by 
$\sim$20kms$^{-1}$) over $\sim$1Gyr (Asian, Figueras \& Torra 1999). Moving groups younger than 
$\sim$1Gyr should thus consist of young populations with characteristic space motions, and membership 
of such a group will accurately constrain the age of an eclipsing BD-M dwarf binary just as it 
would do for open cluster members.

The most notable such groups in the solar neighbourhood are listed in Table 2 along with their 
kinematic motions and age. Most have open cluster(s) associated with them, which range in size from 
the rich Pleiades cluster down to the almost evaporated Ursa-Major cluster. In general, each group 
has one (or two in the case of the Pleiades) distinct kinematic signature. However, note that the 
Gould belt is rather more complex than the others, since it consists of an agglomeration of several 
very young kinematic groups superposed on a less distinct velocity ellipsoid (Moreno, Alfaro \& Franco, 
1999). Each of the distinct kinematic sub-groups is believed to have a common origin (eg. Asian et~al., 
1999), with the ellipsoid possibly resulting from an expanding ring of star formation (the local bubble). 
Interestingly, it appears that a large fraction of young local stars reside in these moving groups. 
Dehnen (1998) has shown (using Hipparcos astrometry) that the majority of colour selected young disk 
stars are moving group members, and Makarov (2003) found that a high fraction of X-ray active stars 
within 50pc belong to the Pleiades and Gould Belt moving groups.

\subsection{Identifying young potential moving group members}

In order to separate possible young moving group members from the large population of older disk stars, 
one needs to observationally identify young sources. Age constraints can be placed on young late K and 
early M dwarfs in a variety of ways. Chromospheric/coronal H$_{\alpha}$ emission from open cluster members 
(Hawley et~al. 1999; Gizis, Reid \& Hawley 2002) have shown that there is a well defined, age dependent 
colour beyond which activity becomes ubiquitous. Despite considerable scatter, all stars redder than 
this colour are dMe (EW$_{H\alpha}\ge 1.0$\AA), while the bluer stars are dM without emission. This so 
called ``H$_{\alpha}$ limit'' colour increases with the age of the population, and can identify (in 
emission) K5-M0 dwarfs younger than $\sim$30Myrs, M0-3 dwarfs younger than 100Myrs, and M3-5 dwarfs 
younger than $\sim$600Myrs. Lithium abundances can also provide age constraints. Late K dwarfs deplete 
lithium over $\sim$100Myrs, and early-mid M dwarfs deplete lithium over $\sim$30-50Myrs (Preibisch et~al. 
2001; Zapatero-Osorio et~al. 2002; Jeffries et~al. 2003; Randich et~al 2000). One may also constrain ages 
by fitting isochrones to colour-magnitude data. A $\sim$0.5M$_{\odot}$ star will not have fully contracted 
until it is $\sim$100Myrs old. Therefore, with an accurate parallax distance (possible even at 500pc using 
the VLT as an interferometer), an appropriate colour-magnitude diagram (see Stauffer et~al. 2003), and 
photometric corrections to account for metallicity sensitive colours (eg. Kotoneva, Flynn \& Jimenez 
2002), the age of such a star could be approximately constrained.

Clearly, identifying late K and early M dwarfs younger than $\sim$100Myrs will not be difficult. However, 
older counterparts will be fully contracted, show no lithium and have H$_{\alpha}$ emission $<$1\AA. Such 
sources will require higher resolution analysis if their ages are to be constrained from their activity levels. 
For instance, one could gauge activity levels by measuring Mg II h \& k and Ca II H and K (Doyle et al. 1994; 
Christian et al 2001; Mathioudakis \& Doyle 1991), or by measuring H$_{\alpha}$ at high resolution, and 
look for weaker line emission in the core of the absorption line (eg. Short \& Doyle 1998). These methods 
allow one to study activity levels down to those associated with old disk M dwarfs (see Doyle et~al. 1994), 
and should therefore provide an effective way to separate young ($<$1Gyr) late K and early M dwarfs from 
older disk stars.

\subsection{Kinematic membership of moving groups}

Establishing kinematic membership of moving groups will require space motions accurate to a few kms$^{-1}$. 
Since accurate radial velocity curves will have been previously measured (to determine BD mass; see Section 
4.2), these will provide very accurate centre of mass radial velocities for the binaries. Although it should 
be possible to derive some proper motions (with the required accuracy) from existing archive measurements 
(eg. with SuperCOSMOS) sources with distances near 500pc would require some additional effort. Proper motion 
accuracies would need to be at the milli-arcsecond/year level (cf. 5kms$^{-1}\sim$2mas/yr at 500pc), and it 
is now quite feasible to measure proper motions of this accuracy using adaptive optics technology with a 
baseline of $\sim$1-2 years (cf. 3mas positional accuracy; Close et al.., 2003).

The confidence with which one can assign kinematic membership of a particular moving group will depend on 
how unique the group's measured kinematic signature is compared to other young moving groups and any non 
moving group population of young stars (see Figures 1 and 2 of Montes et~al. 2001). The Hyades moving group 
for instance has a space motion well separated from the local standard of rest, but has a rather larger 
internal dispersion compared to some younger moving groups. As a result some Hyades moving group members will 
not be obvious kinematic members of the group. At the other extreme, the very young members of the Gould 
belt and the Pleiades moving group components will not show as much internal dispersion (within each 
component). However, due to their similar space motions, there is some kinematic overlap between some of 
these components. Furthermore, young stars that are not members of any moving group can act as a source 
of potential contamination, if they have the same space motion as the group. However, if the majority of 
young stars are moving group members (see Section 5.2), such contamination should not be at a high level, 
and will only decrease membership probabilities by a small amount.

While it is clear that only some fraction of moving group members will be easily identifiable from their 
kinematics, it is by no means clear exactly what this fraction will be, since there are insufficient radial 
velocity measurements of young disk stars to properly characterise our expectations. However, it is worth 
noting that moving group characterisation will certainly improve in the near future, as large scale radial 
velocity surveys (such as RaVE; Steinmetz 2003) provide radial velocity and metallicity for complete 
magnitude limited samples.

\section{Simulating the number of useful eclipsing BD companions}
We will now describe a simulation we have made to predict the size and properties of the population of 
fully eclipsing BD companions that could be identified by the SuperWASP survey over the whole sky. This 
simulation predicts how many BDs will be identified, how many of these will be detectable from their 
secondary eclipse, how many are likely to be significantly irradiated by their primary, and how many 
will be young (and thus potentially moving group members).

\subsection{The simulation}
For our simulation, we consider low-mass stars with BD companions in close ($\sim$0.02AU) orbits. These 
binaries will have short orbital periods of 1--1.5d. We thus expect $\sim$40-60 transits/binary during a $\sim$2 
month SuperWASP monitoring period, $\sim$11--16 of which should occur while SuperWASP is taking data (allowing 
for typical 9h nights and $\sim$25\% time lost to bad weather). SuperWASP should experience $\sim$9.5 dark 
nights per lunation, and the moon will be set for typically $\sim$20\% of non-dark nights. We therefore expect 
dark sky conditions $\sim$45\% of the time, and would expect $\sim$5--7 transits to occur during this time. It 
is therefore appropriate to use the dark sky distance limits for transit detection. We also assumed that 
for late K dwarfs (V$\sim$16 near the SuperWASP distance limits) 30\% and 75\% will be lost due to blending 
for b$>$20$^{\circ}$ and b$<$20$^{\circ}$ respectively. For the early M dwarfs (V$\sim$18), $\sim$50\% and 
85\% will be blended for b$>$20$^{\circ}$ and b$<$20$^{\circ}$ respectively (see Section 3.4). We therefore 
chose to consider only regions of sky with b$>$20$^{\circ}$, and combined this sky coverage with our distance 
sensitivities to derived space volumes. We then combine these volumes with the nearby luminosity function of 
Reid Gizis \& Hawley (2002) and the vertical disk density law from Zheng et~al. (2001), to derived the number 
of stars within the transit detection limits as a function of M$_V$. We then factored in a 1\% BD companion 
fraction at a separation of 0.02AU for late K and M dwarfs (M$_V\ge$8), with a negligible binary fraction 
for earlier stars (see Section 2.1), and thus determined the fraction of binaries that should be fully 
eclipsing using the equations from Section 2.2.

In order to factor in how many of these BDs will be detectable from the secondary eclipse, we simulated 
the BD T$_{eff}$ distribution. This was done by defining an array of $\sim$8000 BDs (0.015-0.075M$_{\odot}$) 
where we assumed a population mass function with $\alpha$=0.5, and a uniformly distributed random age 
distribution from 0--10Gyrs. For each BD we then used the NextGen and DUSTY (Chabrier et al. 2000) model 
isochrones to estimate T$_{eff}$ from mass and age, using linear interpolation between isochrone grid 
points. We chose NextGen T$_{eff}$s higher than 2500K, and DUSTY T$_{eff}$s where the NextGen T$_{eff}$s 
were $<$2500K. The resulting histogram of T$_{eff}$s is shown in Figure 7. Spectral type ranges (some shown 
on the figure) were determined using the spectral type -- T$_{eff}$ relations from Reid et al. (1999; M6-L2), 
and Knapp et al. (2004: for $>$L2). For each of our selected spectral type ranges (M6$\pm$1-T4$\pm$1) we added 
up all the BDs in the appropriate T$_{eff}$ range, and divided by the total number, so as to represent the 
fraction of BDs in that range. These results are tabulated in Figure 7. The implications of this T$_{eff}$ 
distribution will be discussed in the next section. Finally, we removed any binaries in which the primary was 
too bright compared to the BD companion (using the M$_V$ limits determined in Section 4.1).

\subsection{Simulation results}
The simulation predicts that a total of $\sim$1000 eclipsing BD companions could be discovered by SuperWASP 
over the full La Palma sky. Of these, secondary eclipse measurements should allow the BD brightness to 
be measured (to 10\% accuracy) for $\sim$150 systems. The predicted distribution of primary and BD 
spectral types, and BD mass and age for these 150 systems is shown in Figure 8. Note that these predicted 
results scale with the assumed BD binary fraction and 1/$a$, and can thus be easily scaled up or down for 
different input values. In the top left histogram, the favoured primary spectral types are the earlier 
types, since these stars are brighter, and detectable out to greater distances. In the top right histogram, 
there is a strong preference for a BD spectral class of L4$\pm$1. This preference results from three 
factors in our simulation. Firstly, the spectral type--T$_{eff}$ relation we used (Knapp et al. 2004) 
gives a relatively large T$_{eff}$ range for the L3-5 spectral classes (1500-1900K; see Figure 7). Secondly, 
L4 companions will be detectable around all the primaries that our simulation considers as potential hosts 
for BD companions (M$_V>$8; see Figure 6). Thirdly, although we assume an approximate BD upper mass 
limit of 0.075M$_{\odot}$, the Dusty models predict a m$_{HBMM}\sim$0.07. Our simulation thus has the 
higher mass (0.07-0.075M$_{\odot}$) objects stabilising their T$_{eff}$s around 1500-1900K (ie. L4$\pm$1) 
in the 1--10Gyr age range. Uncertainties in the spectral type--T$_{eff}$ relation and the precise value 
of m$_{HBMM}$ could thus change the spectral type and width of the peak somewhat. However, note that we 
certainly expect an L4 predominance over later companions, since the L4s are detectable around the brighter 
primaries. The bottom left histogram shows this preference for 0.070--0.075M$_{\odot}$ objects. Finally, 
the bottom right histogram shows a strong preference for young BDs, because young BDs are brighter and 
more easily detectable from their secondary eclipses. The age histogram has $\sim$20 binaries younger 
than 1Gyr, $\sim$10 with ages 1-2Gyrs, and a continuum of $\sim$2/Gyr for 2-10Gyrs. This relatively flat 
age continuum is mainly populated by the L4$\pm$1 type companions.

But how many of these detectable BDs would be significantly irradiated? It can be seen by looking at 
Figure 2 that no BDs with spectral types $\le$L0 would be significantly irradiated by any of the primaries 
we consider. However, an L4 BD would be slightly ($\le$25\%) irradiated by the M$_V$=8-9 primaries. Later 
BDs would be irradiated by steadily fainter primaries, and the small number of T5 BDs would be irradiated 
by the majority of primaries. We estimate, using Figure 2 and the top two histograms in Figure 8, that 
$\sim$70\% of the predicted companion BDs will not be irradiated by more than 10\%.

\subsection{A subset of young systems}
We also wish to consider the properties of a young subset of the BDs. Such systems are preferred by our 
detection criteria because BDs are brighter when young. As we discussed in Section 5, many young systems 
should be members of kinematic moving groups, and as such have ages that can be well constrained. We 
therefore re-ran our simulation using a T$_{eff}$ distribution derived by only selecting BDs with ages 
$\le$1Gyr. This simulation predicted $\sim$20 binary systems in which BDs could be well measured from the 
secondary eclipse. The histograms describing these binaries are shown in Figure 9. The primary spectral 
type histogram is similar in shape to that of the main simulation. The BD spectral type histogram however, 
is somewhat different to the main simulation. The L4 peak is still there, but has been significantly 
reduced in number. This is because although the L3--5 T$_{eff}$ range is still relatively large, the 
majority of the 0.070--0.075M$_{\odot}$ companions (ie. in the 1-10Gyr age range) have been excluded in 
this age limited simulation. However, there is now an additional peak for M5--9 BDs. This peak results 
from the fact that all M5--9 dwarfs must be young in order to be BDs, and will thus not be removed by our 
age constraint. The mass histogram still shows a general preference for higher mass BDs, but the removal 
of the large number of older 0.070--0.075M$_{\odot}$ objects has decreased the highest mass point 
significantly. The constrained age histogram is shown with a finer scale, and the preference for younger 
systems clearly remains. As before, we have used Figure 2 and the top two histograms (this time from 
Figure 9) to estimate irradiation effects, and estimate that $\sim$80\% of the predicted young BD 
companions will not be irradiated by more than 10\%. The fraction of the $\sim$16 binaries (containing 
un-irradiated BD companions) that will be clearly associated with a moving group is unclear (see Section 
5.4). However, it seems likely that we might expect $\sim$10 systems capable of providing BD mass, age, 
radius, metallicity and luminosity data, for our assumed 1\% BD companion fraction.
\section{Identification and followup}
\subsection{Detection rate}
On-line optical survey data suggests that we would expect $\sim$300,000--1 million sources to 
V$\sim$18 in a 320 sq. deg area of sky with $|b|>20^{\circ}$ (ie. a typical SuperWASP-I pointing). 
Typically $\sim$70\% of these will have V=16--18, and $\sim$20\% V=14--16. About 20\% of sources 
will be late K and early M dwarfs ($\sim$100,000 per 320 sq. degs), with the majority being earlier 
G and K dwarfs. Of these, $\sim$20,000 will be monitored with a sufficient accuracy to detect BD 
transits (ie. mostly in the V=14--16 range). Typically $\sim$30\% ($\sim$6000) of these are expected 
to be slightly blended (at the $\ge$1\% flux level) with neighbouring sources. We do not expect 
random fluctuations to produce more than $\sim$1 false detection of a 5-$\sigma$ transit event in 
every $\sim$2 million sources (ie. in the full sky late K and M dwarf population). False positives 
will more likely result from unresolved eclipsing stellar systems in which the eclipsing binary 
component contributes only a small fraction of the total light. Alonso et al. (2004) report the 
discovery of 1 transiting planet out of 16 candidates (where the false positives were dominated 
by such sources).

We might expect $\sim$5 genuine fully eclipsing BD companions to be discovered in each $\sim$320 sq. 
deg area of sky (or for instance, 30 per year from SuperWASP-I). Of these, we would expect $\sim$15\% 
of binaries ($\sim$5 per year) to have measurable (10\% accuracy) secondary eclipses, and $\sim$1/3 
of these to have ages $<$1Gyr (ie. 1--2 per year). These BD detection rates may be scaled for alternative 
binary criteria -- they will scale with BD companion fraction and 1/$a$. Note also that the actual 
transit candidate numbers will be higher, since we have only considered full eclipses, and also expect 
additional transits by faint low-mass stellar companions, and possibly planetary companions to pass our 
selection criteria. However, the level at which these will occur is beyond the scope of this paper.

\subsection{Followup procedures}
The task of identifying eclipsing BD companions will begin with a search of the transit survey database. 
Automated search algorithms should be used to identify transit signatures in the light curves, allowing 
for the possibility that there could also be some variation in brightness due to rotationally modulated 
spot coverage. Large pixels may result in the blending of some sources, so it will be important to confirm the 
transit source at better spatial resolution. Optical (eg. from the STSci Digitized Sky Survey, or the 
Super Cosmos Sky Survey) and near infrared (from 2MASS) sky survey images will show if a transit source 
is a blend of several sources or a single star, and also provide optical/NIR colours and basic 
astrometry. These data should also differentiate between dwarfs and giants, confirming or otherwise the 
faint nature of a transiting companion. It may well also be desirable at this stage to measure additional 
optical light curve data (using a larger telescope), to both confirm the transit source, and more accurately 
measure the primary eclipse profile. With the addition of a measured spectral type (and metallicity), 
the character of the transit will have been well established.

Having confirmed a low luminosity eclipsing companion of a cool dwarf, the next step will be to attempt 
to measure the secondary eclipse profile and obtain high resolution radial velocity data (see Sections 
4.1 and 4.2). A parallax distance to the primary will also be required to confirm its brightness (from 
which the BD brightness is inferred). Together these data could provide BD mass, radius, metallicity and 
brightness. The final stage will be to try and determine the age of the system, by measuring accurate 
astrometry and considering all relevant moving group membership criteria (see Section 5.3).

\subsection{Additional issues for young sources}
Many young stars may show periodic photometric variability resulting from rotationally modulated 
spot coverage. This generally occurs at the 5-10\% level (Krishnamurthi etal 1998), has periods 
of $\sim$1-5 days (Terndrup et~al 2000), and maintains the same modulation shape over numerous 
rotation periods (Young et~al. 1990). The binary systems we consider should produce primary eclipses 
at least 4\% deep, lasting for $\sim$1.5hrs, with a period of $\sim$1.5d. With such distinct forms of 
variation, automated search algorithms should be able to identify transit sources even if they show 
some spot modulation.

Accurately measuring the much fainter secondary eclipses from followup NIR light curves of variable 
sources could prove more of a challenge. If the primary rotation and BD orbital periods are different, 
one could simply determine the form of the rotational modulation by phasing the data appropriately, 
and averaging over several rotation periods. The rotational modulation could then be subtracted off 
the light curves, leaving only the secondary eclipse dips. However, if the rotation and orbital periods 
are the same (close binaries can become phase locked by tidal forces; see below), then the eclipse 
times will always correspond to the same rotational phase. 

For a 0.05M$_{\odot}$ BD at 0.02AU separation from a primary with mass from 0.6--0.35M$_{\odot}$ 
(late K -- early M), the spin-orbit synchronisation time-scale varies from $\sim$0.04--2.5Gyrs 
respectively (depending primarily on the primary radius and companion orbital separation; Udry et~al. 
2002). Such close BD companions around late K dwarfs are thus likely to be in synchronised orbits, 
but those around early M dwarfs will probably not have had time to synchronise. To deal with synchronised 
sources, one could measure simultaneous R-, I- and K-band light curves. The optical bands would contain 
a negligible fraction of BD flux, and any variability would thus be purely due to rotational modulation. 
The colour-colour relationship between R-K and R-I modulation away from eclipse could be well determined 
from the light curve data. This relationship will depend on how the flux in each band changes as spots 
rotate onto, around and off the observable face of the star. This will of course depend on the temperature, 
size and surface distribution of spots, but should be a tight relationship for any particular spot 
configuration. This colour-colour relationship could thus be used to convert R-I measurements (taken 
during eclipse) into R-K measurements (as if the BD was not eclipsing). The R-band light curve could 
then be converted into a ``pseudo-K-band light curve'' (using this R-K colour), which would show only 
the K-band rotational modulation during eclipse. This could then be subtracted off the original light 
curve as before. Simultaneous optical/NIR observations would be best done with a robotic telescope with 
infrared/optical capabilities. The 2-m robotic Liverpool Telescope for example could measure an 
appropriate set of R- I- and K-band magnitudes in $\sim$25 minutes. This would then require light 
curve coverage of more transits to obtain the required accuracy.

The radial velocity data could also be affected by rotationally modulated spot coverage. As photospheric 
dark spots and brighter regions rotate over the surface of the star, they can produce radial 
velocity modulation. These rotational radial velocity variations will have an amplitude up to v$\sin{i}$, 
a period equal to the rotation period (P$_{rot}$), and a oscillatory shape that depends on the distribution 
of spots across the stellar surface. For illustration, early M stars with ages $<$100Myrs fall into three 
categories of rotation rate. The majority ($\sim$60-70\%) have slower v$\sin{i}<$20kms$^{-1}$ rotation 
velocities (with rotation periods of $\sim$1-5days). $\sim$10-20\% are faster rotators (v$\sin{i}=$30-60
kms$^{-1}$), and only 5-20\% are ultra-fast rotators, with v$\sin{i}>$100kms$^{-1}$ (see Mochnacki et~al. 
2002, and Terndrup et~al, 2000, for young field stars and Pleiades stars respectively). The radial velocity 
oscillation we expect close BD companions to induce is $\sim\pm$20kms$^{-1}$, with a period of $\sim$1.5days. 
Clearly in the majority of cases, the rotational velocity modulation will be smaller than the orbital 
modulation induced by the BD. In general, our task will simply be to deconvolve the two radial velocity 
components. This should generally be easy. For synchronised orbits, one could measure additional radial 
velocity data at a later date, when the spot modulation has changed.

\section{Conclusions}
In order to assess the prospects of discovering eclipsing BD companions capable of yielding 
mass, age, radius, metallicity and luminosity data, we have assessed what is currently known about such 
systems. Although close BD companions are fairly rare around late K and early-mid M dwarfs, there could 
be a $\sim$1\% companion fraction at very close ($\sim$0.01--0.02AU) separation. Our knowledge of this 
companion fraction is poor, but if correct, it suggests that although the BD desert extends into this 
regime, it is not as dry as for solar type stars.

Using a simulation, and the SuperWASP project as an example, we have predicted the numbers and 
properties of close BD companions that could be identified by ultra-wide-field transit surveys in the 
next few years. We assumed a 1\% close BD companion fraction at 0.02AU, a uniform age range from 0-10Gyrs, 
and a BD mass function of $\alpha$=0.5. We estimate SuperWASP to be sensitivite to transits around $\sim$1 
million late K/early-mid M dwarfs. Based on our current knowledge of BD properties our simulation predicts 
that $\sim$400 BDs in eclipsing binaries could be found. It predicts that $\sim$60 of these binaries 
should have secondary eclipses that are measurable from followup NIR light curves, and should thus 
provide direct measurement of the BD brightness. The most common detectable BDs are expected to be 
higher mass BDs orbiting late K-M3 primaries. The majority ($\sim$70\%) of the detectable BDs should 
not be irradiated by $>$10\%.

The simulation also reveals a preference for youth amongst binaries containing detectable BDs. This 
results from the fact that younger BDs are brighter relative to their primary. Our simulation predicts 
that $\sim$20 binaries containing detectable BDs will have ages $<$1Gyr. We suggest that it is likely 
that a significant fraction of such young systems could be young moving group members - disk populations 
with characteristic space motions that are believed to share a common origin, and thus have an age that 
can be well constrained (see Section 5). The typical spectral type of the BDs in this young sub-sample 
ranges from M5--L5, and may be bi-modal, peaking at M5--9 and L3--5. Higher mass BDs are still preferred, 
but the preference is not as strong for the young subset, and the population is more evenly spread 
over the mass range 0.040-0.075M$_{\odot}$. We expect $\sim$80\% of these young companion BDs not to 
be irradiated by $>$10\%.

Although some of our input assumptions are rather uncertain, we show that transit surveys such as 
SuperWASP could find significant numbers of eclipsing BD companions, and by carrying out the followup 
measurements we describe, it would be possible to accurately determine mass, age, radius, metallicity, 
and luminosity data for $\sim$10--20 young ($<$1Gyr age) BDs. The majority ($\sim$80\%) of these should 
not be irradiated above the 10\% level, and could thus provide an empirical testbed to compare to 
theoretical models of free-floating BDs.

\acknowledgments
We thank Don Pollacco, Francis Keenan, Pete Wheatley, Richard West \& Simon Hodgkin for useful 
discussions. The authors are grateful to PPARC for their support of this work.

\clearpage


\begin{figure}
\epsscale{0.8}
\plotone{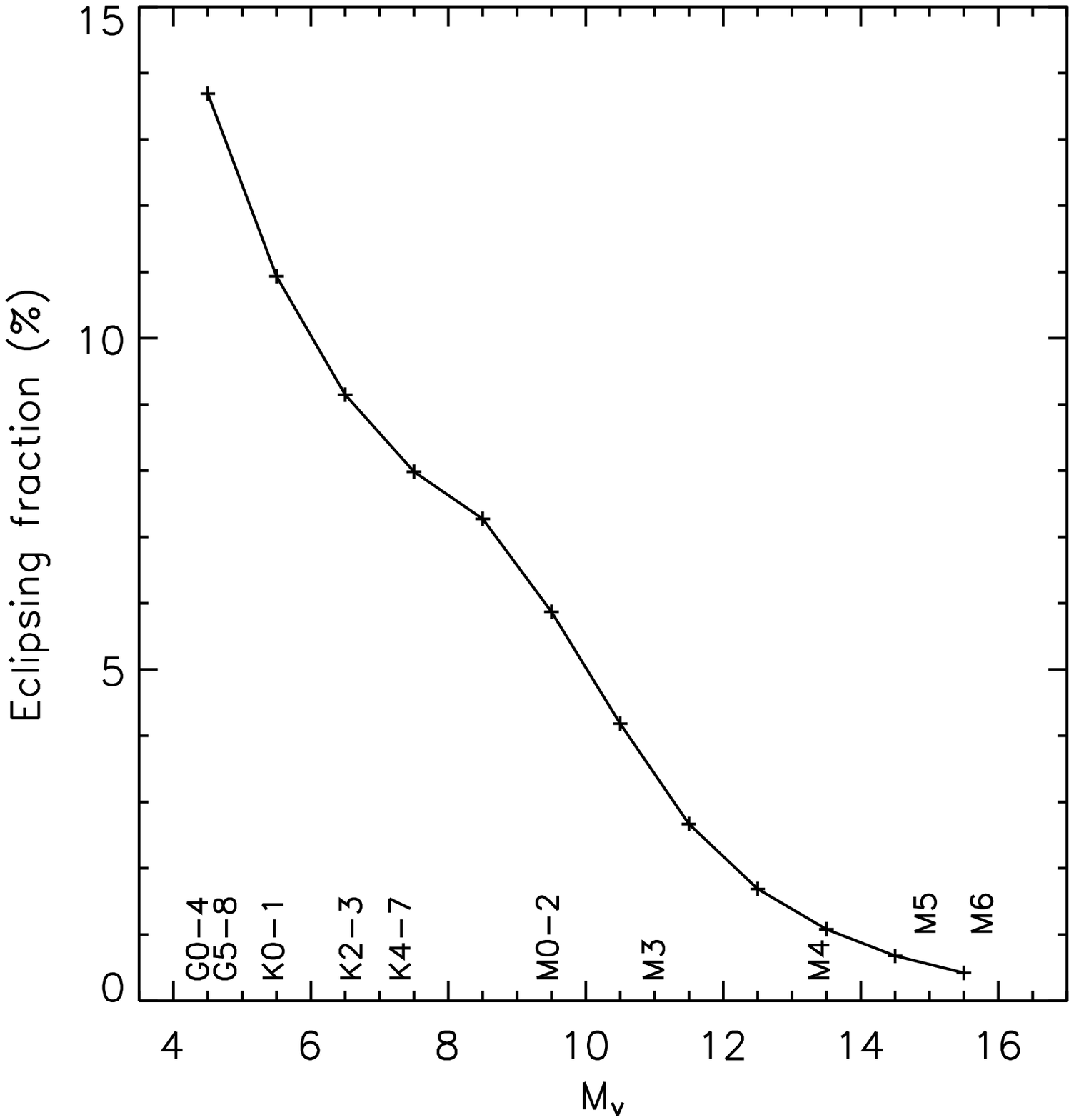}
\caption{The fraction of companion BDs (0.1R$_{\odot}$) that will fully 
eclipse their primary star at a separation of 0.02AU, against primary type.}
\end{figure}
\clearpage

\begin{figure}
\epsscale{0.8}
\plotone{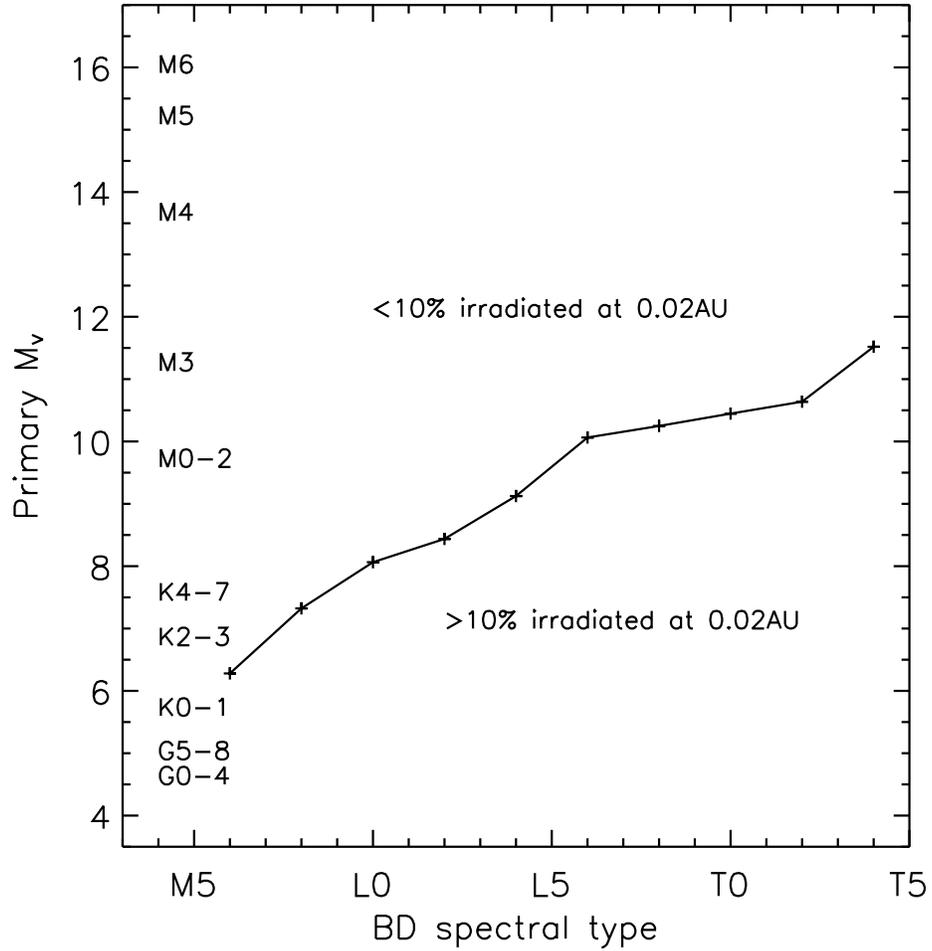}
\caption{Primary M$_V$ against the companion BD spectral type. The solid 
line represents binary systems in which the BDs are irradiated by 10\% at 
a separation of 0.02AU. The regions above and below the line represent 
lower and higher levels of BD irradiation respectively.}
\end{figure}
\clearpage

\begin{figure}
\epsscale{0.8}
\plotone{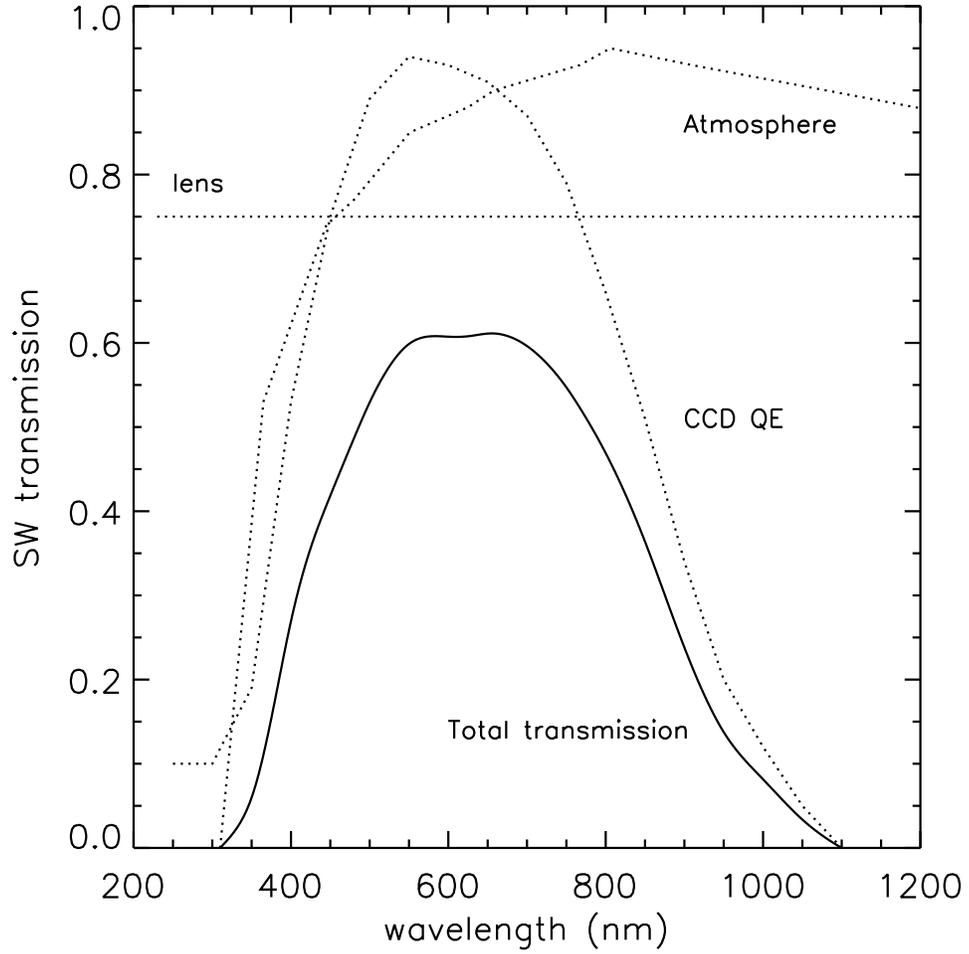}
\caption{The transmission of SuperWASP. Atmospheric transmission 
(airmass=1.5), optical throughput and CCD quantum efficiency are shown with dotted 
lines. The total transmission is shown with a solid line.}
\end{figure}
\clearpage

\begin{figure}
\epsscale{0.8}
\plotone{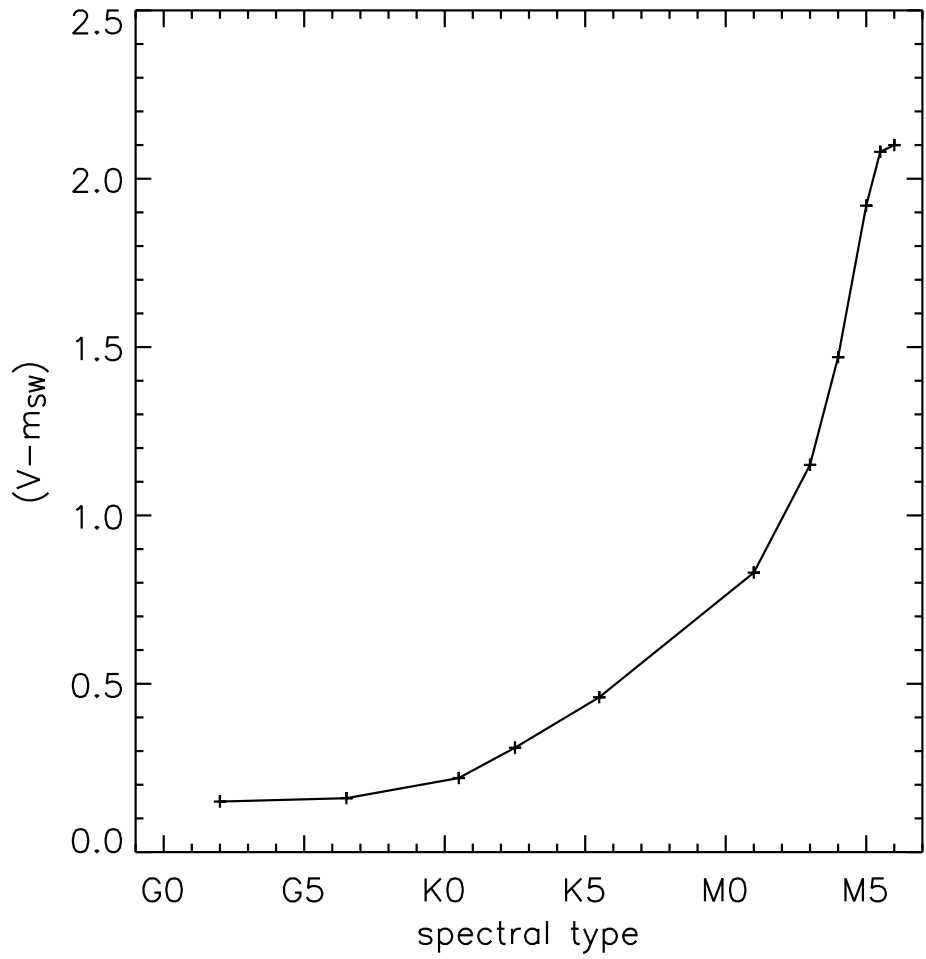}
\caption{Synthetic (V-m$_{SW}$) colour against dwarf spectral type.}
\end{figure}
\clearpage

\begin{figure}
\epsscale{0.8}
\plotone{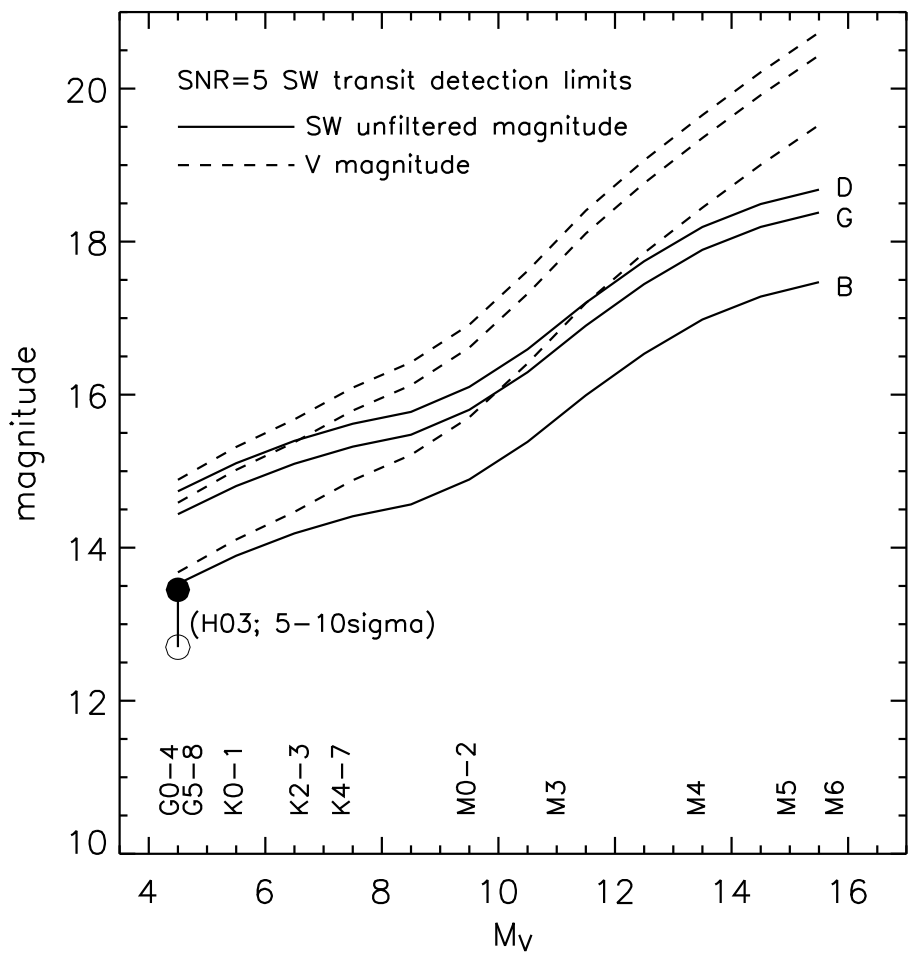}
\caption{SuperWASP magnitude limits for 5-$\sigma$ BD (0.1R$_{\odot}$) transit 
detections against primary type. Limits are shown for dark (D), grey (G) and bright 
(B) skies. Solid lines show magnitudes in the SuperWASP system. Dashed lines signify 
the SuperWASP limits converted into V magnitudes. A point from Horne (2003) is shown 
for comparison.}
\end{figure}
\clearpage

\begin{figure}
\epsscale{0.8}
\plotone{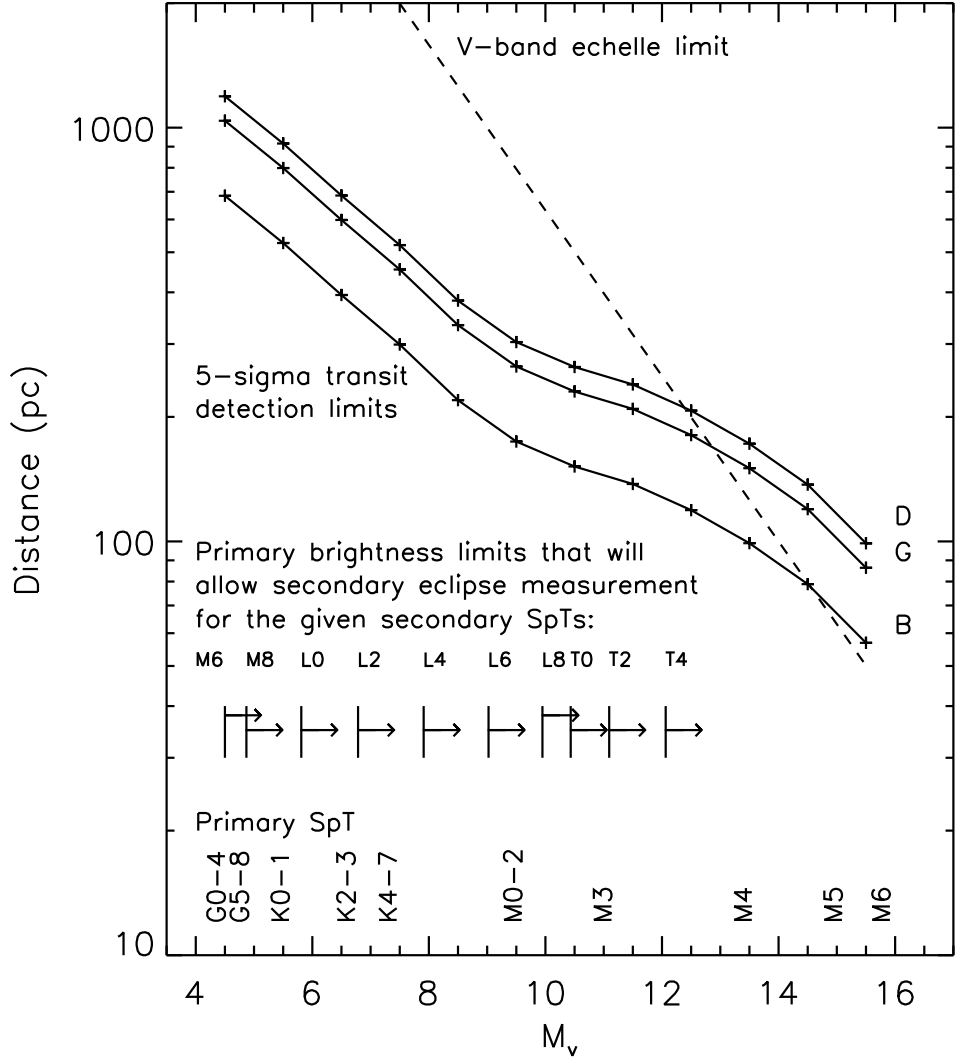}
\caption{Detection and follow-up limits for transiting BD companions. The solid 
lines show the distance limits for 5-$\sigma$ BD transit detections (for dark, 
grey and bright skies) against primary type. The dashed lines indicate distances 
out to which the mass of a 0.05M$_{\odot}$ BD at 0.02AU separation could be 
accurately measured with V- and I-band echelle spectroscopy on an 8-m telescope. 
The M$_V$ lower limits shown indicate the maximum brightness a primary star may 
have if we are to be able to measure the brightness of a BD companion (with spectral 
type of T4-M6) with 10\% accuracy, from a near infrared light curve of the secondary 
eclipse.}
\end{figure}
\clearpage

\begin{figure}
\epsscale{0.8}
\plotone{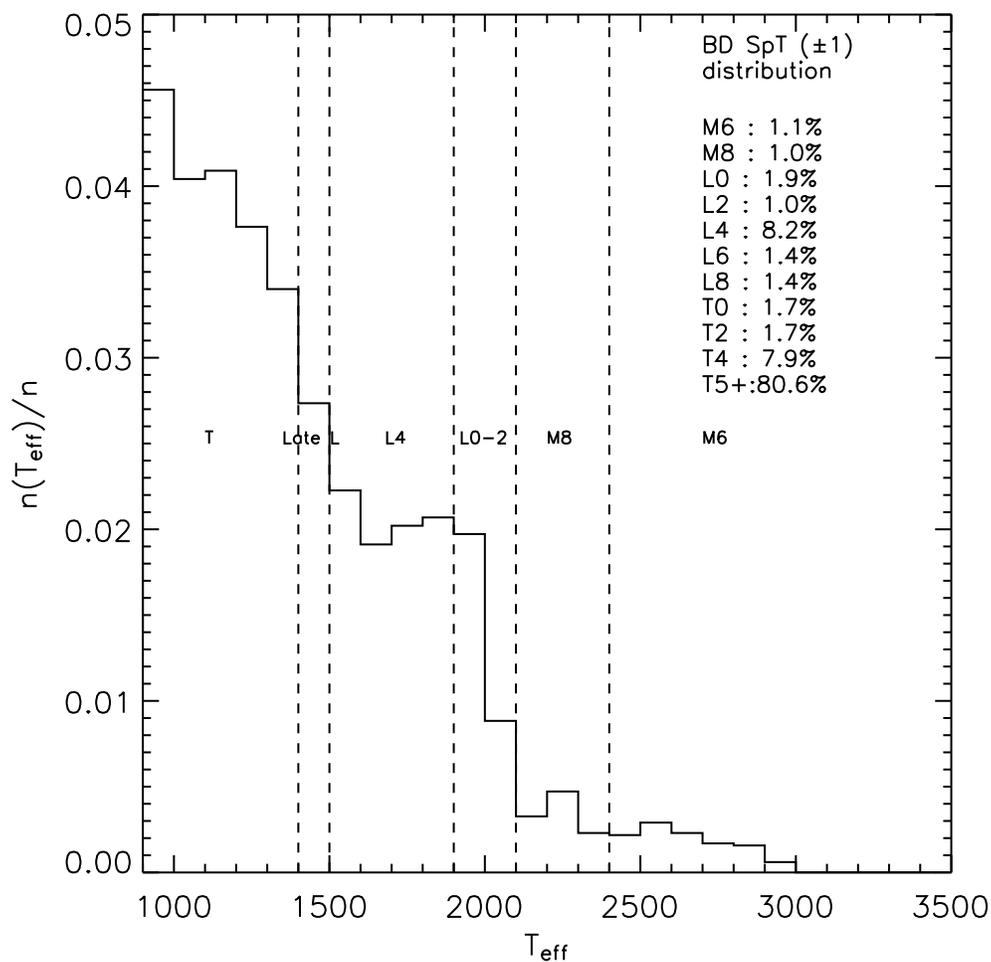}
\caption{The T$_{eff}$ distribution for a population of BDs with a uniform age 
spread from 0-10Gyrs, and an $\alpha$=0.5 mass function. Some spectral type 
divisions are indicated with dashed lines. The fraction of BDs in spectral type 
ranges from M6-T4 is given in the top right of the plot. Note that many BDs are 
cooler than 900K, although the plot does not cover this T$_{eff}$ range.}
\end{figure}
\clearpage

\begin{figure}
\epsscale{1.0}
\plotone{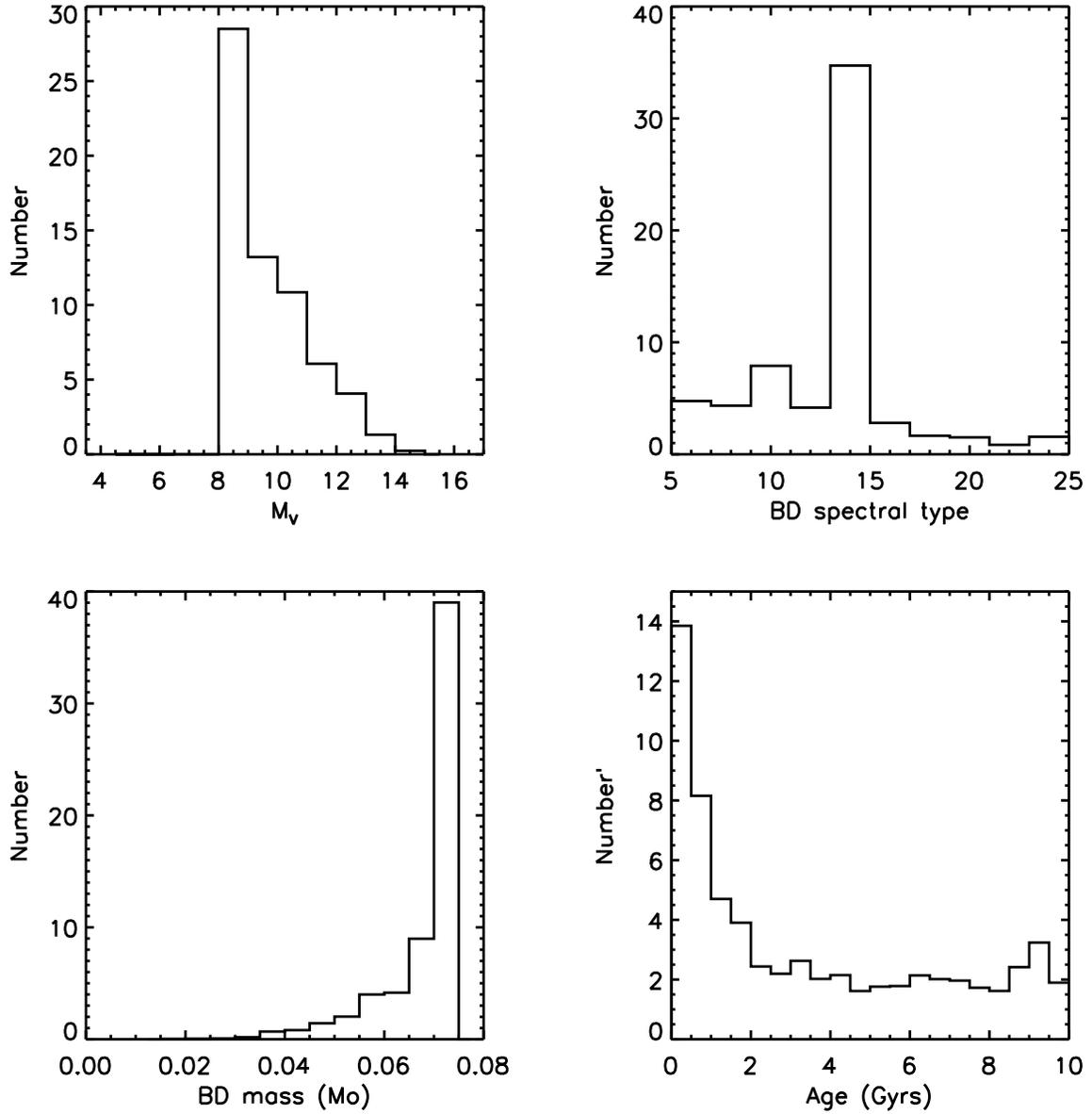}
\caption{Predicted numbers of fully eclipsing brown dwarf companions with 
measurable secondary eclipses, that could be found by SuperWASP. The four 
sub-panels show the primary and BD spectral type distributions, and the 
BD mass and age distributions.}
\end{figure}
\clearpage

\begin{figure}
\epsscale{1.0}
\plotone{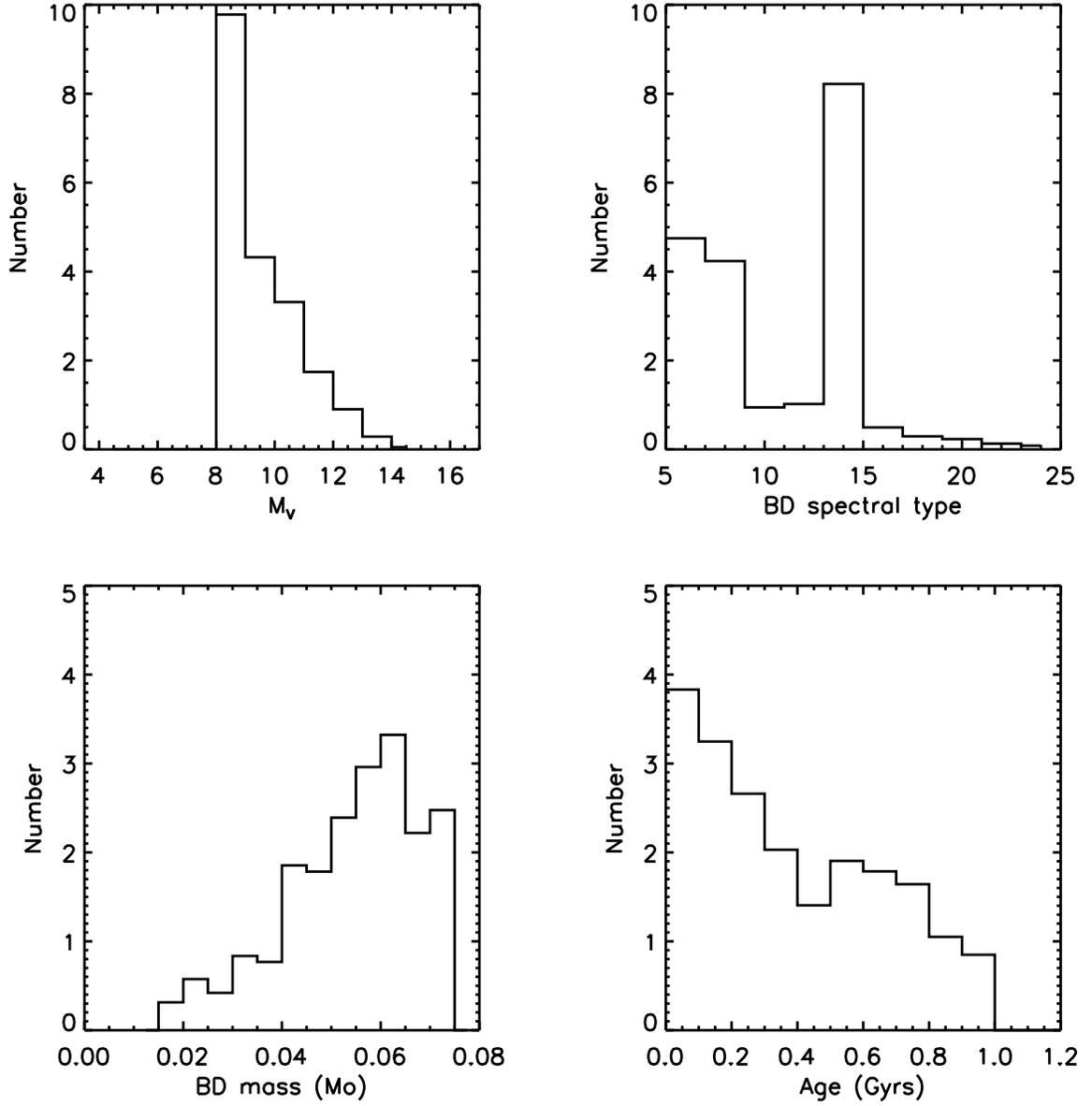}
\caption{The same as figure 8, but for systems with ages $<$1Gyr.}
\end{figure}
\clearpage


\begin{table}
\caption{Brown dwarf companion fractions as a function of primary type. Companion fractions 
have been estimated from the following; (1) Halbwachs et al. (2000); (2) Duquennoy \& Mayor (1991); 
(3) Vogt et al. (2002); (4) Blundell et al. (2004); (5) Els et al. (2001); (6) Lui et al. (2002); 
(7) Gizis et al. (2000); (8) Wilson et al. (2001); (9) Reid \& Mahoney (2000); (10) Marchal (2003); 
(11) Nidever et al. (2003); (12) Oppenheimer et al. (2001); (13) Reiners (2004); (14) Pinfield et al. 
(2003); (15) Burgasser et al. (2003); (16) Close et al. (2001); (17) Reid et al. (2001).}
\begin{tabular}{|l|c|c|c|c|c|c|c|}
\hline
\multicolumn{8}{|c|}{Stellar companions}\\
\hline
Separation (AU) & 0.001--0.01 & 0.01--0.1 & 0.1--1 & 1--10 & 10--100 & 100--1000 & 1000--10000 \\
\hline
F, G \& K & -- & \multicolumn{3}{|c|}{$\sim$4\% per decade}   & \multicolumn{3}{|c|}{$\sim$10--15\% per decade} \\
primaries & (R$_{\odot}$=0.005AU) & \multicolumn{3}{|c|}{(1)} & \multicolumn{3}{|c|}{(2)} \\
\hline
\hline
\multicolumn{8}{|c|}{Brown dwarf companions}\\
\hline
Separation (AU) & 0.001--0.01 & 0.01--0.1 & 0.1--1 & 1--10 & 10--100 & 100--1000 & 1000--10000 \\
\hline
F, G \& K & -- & \multicolumn{3}{|c|}{$<$0.07\% per decade} & ?     & $\sim$10\% & 5 -- 13\% \\
   & (R$_{\odot}$=0.005AU) & \multicolumn{3}{|c|}{(3,4)}      & (5,6) & (7,8)      & (7)    \\
\hline
Early -- mid M & 0 -- 2\% & ? & 1\% & 1\% & 1 -- 3\% & -- & -- \\
               & (9,10)    &   & (11) & (11) & (12) &    &    \\
\hline
late M, L \& T & \multicolumn{3}{|c|}{$\le$30\%} & 10 -- 20\% & -- & --    & -- \\
               & \multicolumn{3}{|c|}{(13,14)}     & (15,16,17)  &    &    &    \\
\hline
\hline
\multicolumn{8}{|l|}{The BD companion fraction around early-mid M dwarfs for a$\le$0.02AU is 0 -- 2\%}\\
\hline
\end{tabular}
\end{table}

\clearpage

\begin{table}
\caption{Summary of the kinematic moving groups that make up the local young 
disk population.}
\begin{tabular}{|l|c|c|l|}
\hline
Moving group     & U, V, W      & Age    & Includes \\
(comments, refs) & (kms$^{-1}$) & (Myrs) &          \\
\hline\hline
{\bf IC 2391}          & -21, -16,  -9 & 35-55     & IC 2391 cluster \\
(1)                    &               &           & \\ \hline
{\bf Gould Belt}       & U$\sim$V$\sim$W & 10-90   & Sco-Cen, Cas-Tau \\
Velocity ellipsoid with& $\sim$-20-0   &           & Carina-Vela and \\
some substructure      &               &           & TW Hydra assocs\\
(some structures       &               &           & \\
sometimes associated   &               &           & \\
with the Pleiades MG)  &               &           & \\
(2, 3, 4)              &               &           & \\ \hline
{\bf Pleiades}         &  -9, -26,  -9 & 100-150   & Pleiades, Alpha Per, \\
(aka Local association.&               &           & and NGC 2516 clusters \\
Contains two main      & -17, -22,  -6 & 200-300   & M34 cluster \\
velocity substructures)&               &           & \\
(5, 6)                 &               &           & \\ \hline
{\bf Castor}           & -11,  -8,  -10& $\sim$320 & The A-stars Vega \\
(7)                    &               &           & and Fomalhault \\ \hline
{\bf Coma Berenices}   & -10,  -5,  -8 & 250-400   & Coma Ber cluster \\
(8)                    &               &           & (aka Melotte 111) \\ \hline
{\bf Ursa Major}       & +15,  +1, -11 & 400-600   & Ursa Major cluster \\
(aka Sirius super-cluster) &           &           & \\
(9)                    &               &           & \\ \hline
\end{tabular}
\end{table}

\clearpage

\addtocounter{table}{-1}
\begin{table}
\caption{Continued.}
\begin{tabular}{|l|c|c|l|}
\hline
{\bf NGC 1901}         & -25, -10, -15 & $\sim$500 & NGC 1901 cluster \\
(10)                    &               &           & \\ \hline
{\bf Hyades}           & -40, -18,  -2 & $\sim$600 & Hyades and Praesepe \\
(11)                   &               &           & clusters \\ \hline
\multicolumn{4}{|l|}{\it References} \\
\multicolumn{4}{|l|}{1. Lynga (1987)} \\
\multicolumn{4}{|l|}{2. Moreno, Alfaro \& Franco (1999)} \\
\multicolumn{4}{|l|}{3. Makarov \& Urban (2000)} \\
\multicolumn{4}{|l|}{4. Song, Zuckerman \& Bessell (2003)} \\
\multicolumn{4}{|l|}{5. Asiain, Figueras \& Torra (1999)} \\
\multicolumn{4}{|l|}{6. Asiain, Figueras, Torra \& Chen (1999)} \\
\multicolumn{4}{|l|}{7. Ribas (2003)} \\
\multicolumn{4}{|l|}{8. Odenkirchen, Soubirian \& Colin (1998)} \\
\multicolumn{4}{|l|}{9. King et~al., 2003} \\
\multicolumn{4}{|l|}{10. Dehnen (1998)} \\
\multicolumn{4}{|l|}{11. Montes et~al., (2001)} \\
\hline
\end{tabular}
\end{table}
\clearpage

\end{document}